 \pdfoutput=1 

\documentclass[letterpaper, 10 pt, conference]{ieeeconf}  

\IEEEoverridecommandlockouts                              

\usepackage{graphics} 
\usepackage{epsfig} 
\usepackage{amsmath} 
\usepackage{amssymb}  
\usepackage{balance}
\usepackage{url}
\usepackage{cite}
\usepackage{subfigure}
 \usepackage{graphicx}
\usepackage{textcomp}
\usepackage{xcolor}
\usepackage{bm}
\usepackage{algorithm}
\usepackage{algpseudocode}

\title{\LARGE \bf
Emergent Structure in Multi-agent Systems Using Geometric Embeddings
}

\author{Dimitria Silveria$^{1}$, Kleber Cabral$^{2}$, Peter Jardine$^{3}$, and Sidney Givigi$^{2}$
\thanks{$^{1}$D. Silveria is with the Department of Electrical and Computer Engineering, 
        Queen's University, 19 Union St, Kingston, ON K7L 3N9
        {\tt\small dimitria.s@queensu.ca}}%
\thanks{$^{2}$  K.\ Cabral and S.\ Givigi are with the School of Computing and the Ingenuity Labs Research Institute, 
        Queen's University, Kingston, ON K7L 3N6 Canada 
        {\tt\small kleber.cabral@queensu.ca,  sidney.givigi@queensu.ca}}%
\thanks{$^{3}$  P. Jardine is with the School of Computing, 
        Queen's University, 19 Union St, Kingston, ON K7L 3N9 
         {\tt\small p.jardine@queens.ca}}%
}

\begin{document}


\maketitle
\thispagestyle{empty}
\pagestyle{empty}

\begin{abstract}
This work investigates the self-organization of multi-agent systems into closed trajectories, a common requirement in unmanned aerial vehicle (UAV) surveillance tasks. In such scenarios, smooth, unbiased control signals save energy and mitigate mechanical strain. We propose a decentralized control system architecture that produces a globally stable emergent structure from local observations only; there is no requirement for agents to share a global plan or follow prescribed trajectories.  Central to our approach is the formulation of an injective virtual embedding induced by rotations from the actual agent positions. This embedding serves as a structure-preserving map around which all agent stabilize their relative positions and permits the use of well-established linear control techniques. We construct the embedding such that it is topologically equivalent to the desired trajectory (i.e., a homeomorphism), thereby preserving the stability characteristics. We demonstrate the versatility of this approach through implementation on a swarm of Quanser QDrone quadcopters. Results demonstrate the quadcopters self-organize into the desired trajectory while maintaining even separation. 

\end{abstract}

\section{Introduction}

Early research in multi-agent flocking demonstrated complex emergent behaviors of practically unlimited scalability can be achieved through the application of simple rules on local observations \cite{reynolds1987}. Using an elegant balance of cohesion, separation, and alignment, Reynolds rules of flocking inspired decades of related research \cite{agronomy13102499}. Such self-organizing and decentralized systems (or, \textit{swarms}) overcome limitations of observability, computation, and communication that typically make large-scale coordination problems challenging to solve. Moreover, swarms of this nature are generally resilient to corruption or loss of individual agents \cite{GAO2024111382}.

\begin{figure}
    \centering
    \includegraphics[width=\columnwidth]{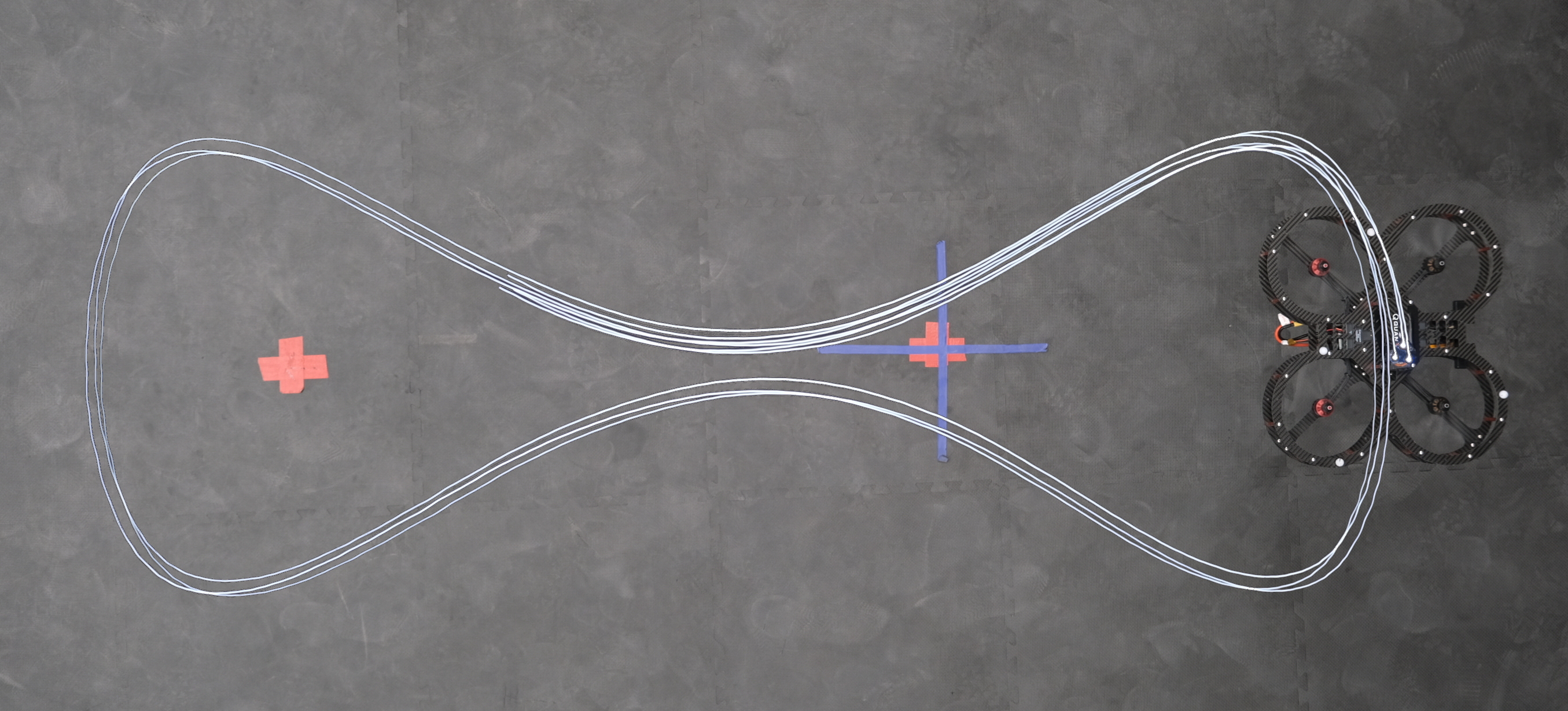}
    \caption{
    The flight path of a UAV forms a dumbbell curve in space, visible from a top-down perspective where the vehicle's curve is traced on the horizontal plane. This trajectory was recorded through the light painting technique, and it tracks the position of LEDs mounted on the top of the UAV. The two red ``X'' on the floor are $1.5$~m distant from each other and can be used as a visual reference. They are not centralized in the trajectory due to the camera perspective.}
    \label{fig:dumbbell-real}
\end{figure}

This work concerns multi-agent systems coordination problems focused on producing stable, fixed geometries composed of the agents themselves. Example of similar work includes lattice formation \cite{olfati2006, HU2022110235}, encirclement \cite{hafez-2015}, dynamic patterns \cite{DONG-2020}, and curved trajectories \cite{fedele-2023}. Specifically, we focus on the dumbbell lemniscatic curve trajectory, which is useful in long-duration unmanned aerial vehicle (UAV) surveillance tasks \cite{Sara-2019, Altan-2020}. Fig.~\ref{fig:dumbbell-real} shows an example of such a trajectory, with one UAV moving through a \textit{dumbbell} curve in our experimental setup.

Various theoretical and algorithmic frameworks have integrated the emergent principles of Reynolds rules with classical control theory. Examples include consensus-based methods \cite{olfati2006}, pinning \cite{WANG2002521}, and distributed control \cite{cao2011}. In \cite{flocking}, the authors propose a decentralized strategy for producing flocking behavior in UAV swarms. In their approach, each agent calculates a local desired velocity based on expected values of relative distance and velocity.
In this scenario, each agent knows its position and velocity with respect to all the other agents in the swarm, which increases the problem's complexity as the number of agents grows. Also, their approach requires the tuning of 11 model parameters for each agent, which requires an additional optimization algorithm. In our approach, each agent only requires knowledge of two adjacent neighbours and selection of three control parameters; this improves scalability and reduces the complexity of large swarms. 

Recent work in \cite{jardine2023} has demonstrated the potential for a new coordination domain based on stabilizing \textit{embeddings}. Embeddings have gained recent attention in the modeling of nonlinear systems. For example, semantic meaning in high-dimensional Large Language Models (LLMs) is commonly captured as word embeddings \cite{NIPS2017_3f5ee243}. Similarly, this concept has been used for input design control of nonlinear dynamic systems \cite{MU2023110673}. 

 Nonlinear control techniques such as robust adaptive control have been used to stabilize swarm trajectories in the presence of uncertainty \cite{adaptive}. However, such approaches are computationally costly for real-time applications. Our technique uses embeddings as a form of feedback linearization. Through a combination of variable changes, state feedback, and quaternion-based rotations, we stabilize a swarm of agents around an embedding using linear control policies. Specifically, we leverage the dynamic encirclement technique from \cite{hafez-2015}. Our embedding is composed of virtual agents induced by quaternion-based rotations from the actual agent positions. Since the quaternion rotations are homeomorphic transformations, the embedding trajectory is topologically equivalent to that of the agents and therefore preserves the desirable stability characteristics. 

While similar embedding-based approaches have been proposed in \cite{jardine2023}, these applications have been constrained to one class of lemniscate trajectory. We expand on this earlier work by introducing more complex curves; specifically, the dumbbell curve. This requires identifying new functional parameters, which we present in a more generalized form, demonstrating that our approach can be extended to a larger set of closed trajectories. 

We also present a new control system architecture that permits the implementation of our technique on real quadcopter dynamics as a high-level trajectory planner. Quadcopters represent one of the most common platforms in commercial applications  \cite{Abdelkader_2021_aerial}. Moreover, our control architecture demonstrates the applicability of this approach to systems with complex low-level control dynamics, such as the nested control architecture used here \cite{Mechali_2021_Distributed}.

The remainder of this paper is organized as follows: Section \ref{sec:problem} formulates the embedding; Section~\ref{sec:control} presents the control architecture; Section~\ref{sec:methodology} describes the implementation of the controller on real UAVs; Section~\ref{sec:results} presents the experimental results; and, finally, Section~\ref{sec:conclusion} contains the conclusions and recommends future work.

\section{Problem Formulation}
\label{sec:problem}

\begin{figure}
    \center
    \subfigure[sub-3D curves][Three-dimensional curve with projections]{
    \includegraphics[width=\columnwidth]{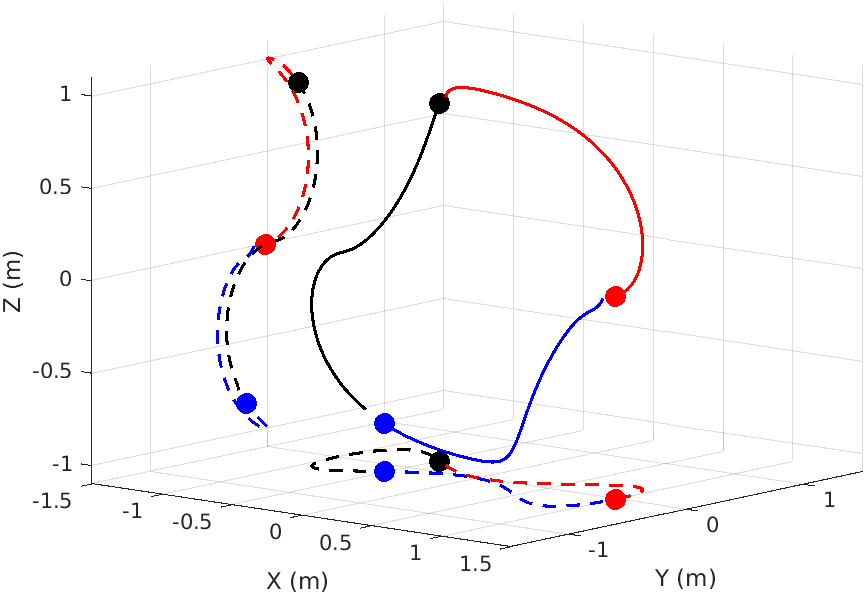} } \\
    \subfigure[sub-dumb-xy][Curve projected in the X-Y plane]{
    \includegraphics[width=\columnwidth]{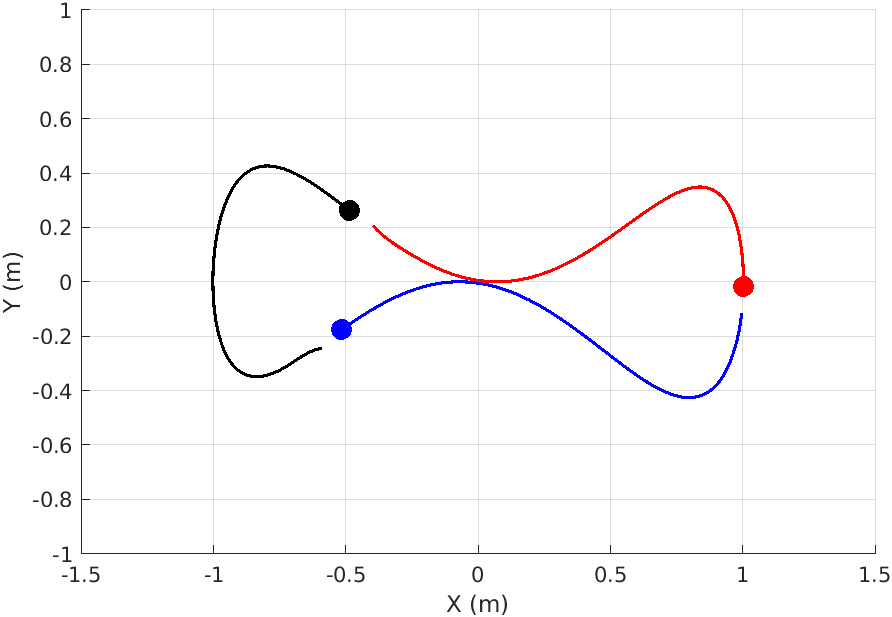}}
    \caption{A \textit{Dumbbell} curve being described by a swarm of three UAVs following the same path. The  with three UAVs following the same path.}
    \label{fig:dumbbell}
\end{figure}

In this section, we formulate a distributed multi-agent control strategy for producing \textit{dumbbell} curve trajectories. Each agent relies on local information to determine its respective position within the larger swarm. The emergent behaviour is characterized by evenly-spaced agents tracking the desired curve in the $x-y$ plane. The number of agents has no theoretical limit, as each agent only requires information about the agent ahead of it (leading agent) and behind it (lagging agent) in the trajectory. In practice, scalability would be constrained by factors such as space, sensor, and communication limitations. 
Fig.\ref{fig:dumbbell}(a) provides an example of such trajectory in 3-dimensions, with Fig.~\ref{fig:dumbbell}(b) illustrating the projection of the dumbbell in the $x-y$ plane. Note that the curve in the $z$ dimension is the result of deforming a circle so that it forms this dumbbell projection. The dots in both figures show the positions of agents. 

Central to the formulation is the concept of a circular embedding, which serves as a structure-preserving map shared by all agents. This embedding is composed of virtual agents induced by rotations from the positions of the actual agents. Inter-agent separation is adjusted with respect to these virtual agents, permitting the use of well-established techniques for dynamic encirclement \cite{hafez-2015}. The overall strategy is based on the lemniscatic arc work of \cite{jardine2023}, but expanded for a broader set of curves. We refer to the resultant structure as a \textit{homeomorphic} polar curve, as it is a topologically equivalent to the circular embedding. 

Our approach relies on quaternion-based rotations. Given angles of rotation $\phi$, $\theta$, $\psi$ around the $x-$, $y-$, and $z-$ axes, respectively, we a construct quaternion rotation vector

\begin{equation} \label{quat_e2q}
\bm{\rho} = 
\begin{bmatrix}
c(\phi/2)c(\theta/2)c(\psi/2) + s(\phi/2) s(\theta/2)s(\psi/2) \\
s(\phi/2)c(\theta/2)c(\psi/2) - c(\phi/2)s(\theta/2)s(\psi/2) \\
c(\phi/2)s(\theta/2)c(\psi/2) + s(\phi/2)c(\theta/2)s(\psi/2) \\
c(\phi/2)c(\theta/2)s(\psi/2) - s(\phi/2)s(\theta/2)c(\psi/2)
\end{bmatrix},
\end{equation}

\noindent where rotation of an arbitrary point in space $\bm{x}=[0, x, y, z]^T$ about the origin is 
\begin{equation} \label{quat_rot}
\bm{x}' = \bm{\rho} * \bm{x} * \bm{\rho}^{-1},
\end{equation}
\noindent where $*$ denotes the Hamilton product \cite{Coxeter-1946}, $\bm{x}'$ is the rotated vector, and $\bm{\rho}^{-1}$ is the quaternion conjugate computed as 
\begin{equation} \label{quat_con}
\bm{\rho}^{-1} =
\begin{bmatrix}
\bm{\rho}[0] &
-\bm{\rho}[1] &
-\bm{\rho}[2] &
-\bm{\rho}[3]
\end{bmatrix}^T,
\end{equation}

\noindent where $\bm\rho[i]$ is the $i$th element of vector $\bm\rho$.


Let us define the parametric equations for a circle of radius $r$ in a horizontal plane around the origin as 
\begin{equation} \label{eqn:circle}
C_c(\phi)
    =
    \begin{bmatrix}
    r\cos{(\phi)}
    \\
    r\sin{(\phi)}
    \end{bmatrix}~\in~\mathbb{R}^2~
    \forall~0\leq\phi<2\pi.
\end{equation}

We further define the parametric equations for a special case of the class of Sextic curves known as the dumbbell curve~\cite{Yates-1974} as
\begin{equation} \label{eq:Dumbell}
C_d(\phi)
= 
\begin{bmatrix}
r\cos{(\phi)} \\
r\sin{(\phi)}\cos^2{(\phi)}
\end{bmatrix}
~\in~\mathbb{R}^2~
    \forall~0\leq\phi<2\pi.
\end{equation}

These types of curves may be useful for various applications including aerial surveillance~\cite{Altan-2020} and long duration operations where smooth, unbiased control signals save energy and mitigate mechanical strain on vehicle components~\cite{Sara-2019}. The dumbbell curve also presents a unique variation on traditional lemniscates, where the elongated approach towards the origin ensures a consistent parallel approach and departure over the origin (as would be required if passing over an aerodrome runway). 
Let us define a functional unit quaternion 
\begin{equation} \label{eq:rho}
\bm{\rho}(\phi) =     
    \begin{bmatrix}
    \alpha(\phi) &
    \beta(\phi) &
    0 &
    0
    \end{bmatrix}^T
    ~\forall~0\leq\phi<2\pi,
\end{equation}

\noindent where 

\begin{align} \label{eqsol3}
\alpha(\phi) &= -\frac{\sqrt{2} \sqrt{\cos(\phi)^2 + 1}}{2}, \\ \label{eqsol3b}
\beta(\phi) &= -\frac{\sqrt{2} \sqrt{-(\cos(\phi) - 1)(\cos(\phi) + 1)}}{2}.
\end{align}

If we express coordinates $(x,y,z)$ as a pure quaternion of the form $\bm{x}=[0~x~y~z]^T~\in~\mathbb{H}^P$, then a dumbbell curve can be produced by mapping ${\bm{x}'}\mapsto \bm\rho(\phi) * \bm{x} *\bm\rho^{-1}(\phi)$. This continuous deformation of a circle is computed as

\begin{equation}
    \begin{bmatrix}
    0 \\
    x' \\
    y' \\
    z'
    \end{bmatrix}
    = 
    \begin{bmatrix}
    \alpha(\phi) \\
    \beta(\phi) \\
    0 \\
    0
    \end{bmatrix}
    *
    \begin{bmatrix}
    0 \\
    r\cos{(\phi)} \\
    r\sin{(\phi)} \\
    0
    \end{bmatrix}
    *
    \begin{bmatrix}
    \alpha(\phi) \\
    -\beta(\phi) \\
    0 \\
    0
    \end{bmatrix}.
\label{eq:rotation}
\end{equation}

Note that the parametric equations for a circle \eqref{eqn:circle} are \textit{embedded} within the larger expression and \eqref{eq:rotation}, which results in the following components

\begin{equation} 
\begin{bmatrix}
x' \\
y'
\end{bmatrix}
=
\begin{bmatrix}
r\cos{(\phi)} \\
r\sin{(\phi)}\cos^2{(\phi)}
\end{bmatrix}
~\forall~0\leq\phi<2\pi,
\end{equation}

\noindent which are the parametric equations for the dumbbell curve in \eqref{eq:Dumbell}. 

The formulation above demonstrates that an agent traveling along a trajectory with states $\bm{x}'$ (i.e., a dumbbell curve) can be controlled with reference to an embedded trajectory with states $\bm{x}$ (i.e., a circle) through a simple transformation. We refer to these embedded states as virtual agents. This formulation is valuable for multi-agent systems coordination, as there are well-established techniques for producing circular trajectories that can now be extended to more elaborate geometries. In this case, the appropriate selection of the functional equations in \ref{eq:rho} provides a map between the coordinates of the dumbbell curve and circle.

\section{Control System Architecture} \label{sec:control}

This section presents a control system architecture for producing the behaviour described in 
Section~\ref{sec:problem} for an arbitrary number of agents. We frame the transformation and change of variables in \eqref{eq:rotation} as a form of feedback linearization, linking our approach to traditional control theory principles. The design is based on double integrator dynamics, which we show later in Section \ref{sec:methodology} is well-suited for extension into real-world applications on autonomous vehicles. 

Let us consider agent $i$ as a particle in free space with position ($\bm{x}_i~\in~\mathbb{R}^3$), velocity ($\bm{v}_i~\in~\mathbb{R}^3$), and inputs ($\bm{u}_i~\in~\mathbb{R}^3$) related by the following dynamics:
\begin{equation} \label{eq:dyn}
\begin{matrix}
\dot{\bm{x}}_i = \bm{v}_i, \\
\dot{\bm{v}}_i = \bm{u}_i,
\end{matrix}
\end{equation}
\noindent where $\bm{x}_i$ is a vector containing Cartesian coordinates and $\bm{u}_i=\ddot{\bm{x}}_i$ is the acceleration of the agent. 


We denote the position of a virtual agent as
 $\hat{\bm{x}}_i$, which is induced by the actual position of vehicle $\bm{x}_i$. These virtual agents form the embedding formulated in Section \ref{sec:problem}. The embedding is an inverse of the rotation in~\eqref{eq:rotation}, essentially \textit{untwisting} the dumbbell curve into a circle, on which we define a linear control policy that stabilizes the induced virtual agent. 
 

Given agent with position and velocity $(\bm{x}(t), \bm{v}(t))$ governed by dynamics \eqref{eq:dyn} over time $t$, quaternion $\bm{\rho}(\phi(t))$ defined in \eqref{eq:rho} and dependent on parameter $\phi(t)$, let us define a virtual vehicle at $\hat{\bm{x}}(t)$ induced by rotation $\hat{\bm{x}}(t)= \bm{\rho}^{-1}(\phi(t)) * \bm{x}(t) * \bm{\rho}(\phi(t))$. The embedding, which acts on $\bm{x}(t)$ through the dynamics \eqref{eq:dyn} to regulate $\hat{\bm{x}}(t)$ is
\begin{equation} \label{eq:equival}
	\lim_{t\to\infty} \hat{\bm{x}}(t)  
	=
	\begin{bmatrix}
	r_d\cos(\dot\phi_dt) \\
	r_d\sin(\dot\phi_dt) \\
	0
	\end{bmatrix},
\end{equation}

\noindent where constant parameters $r_d$ and $\dot\phi_d$ are desired radius and angular speed around the origin. In short, the embedding permits the control of $\bm{x}(t)$ indirectly such that its corresponding virtual vehicle $\hat{\bm{x}}(t)$ is stabilized in a circular trajectory around the origin at radius $r_d$ and $\dot\phi_d$.

When implemented across all agents in the trajectory, this embedding enables the construction of a linear controller that will ensure agents stabilize as follows: 
\begin{equation} \label{eq:rtoInf} 
\begin{matrix}
        \lim_{t\to\infty} \dot{\phi}_{d,i}(t) = \dot{\phi}_{d},  \\ 
        \lim_{t\to\infty} r_{i}(t) = r_{d}.      
\end{matrix}
\end{equation}

\begin{figure*}
    \centering
    \includegraphics[width=\textwidth]{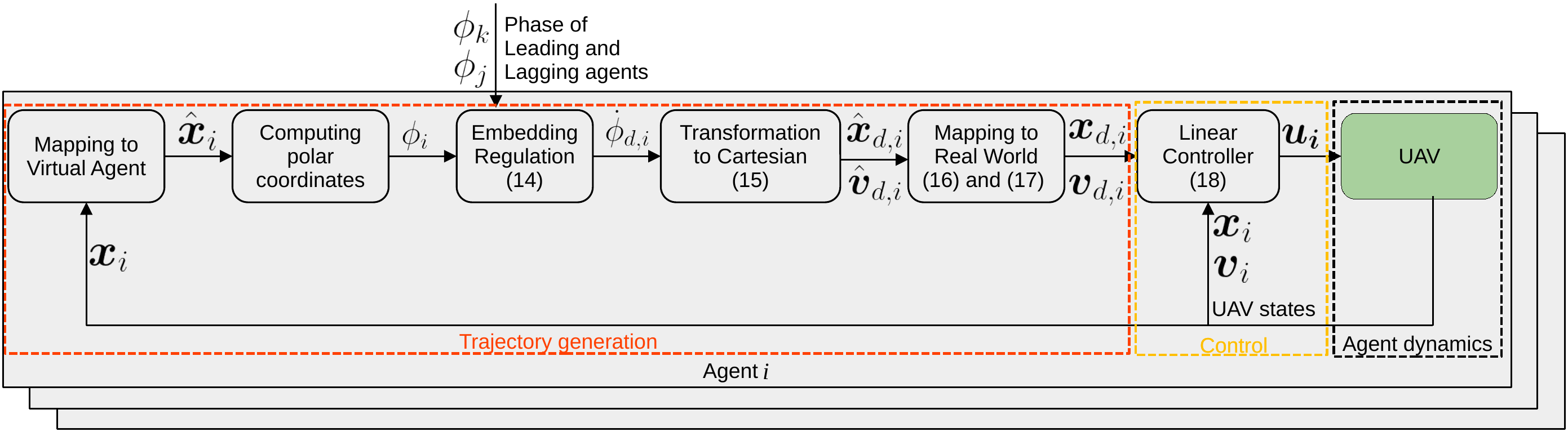}
    \caption{System-level architecture containing the control strategy implemented on each UAV. The numbers in parentheses denote the equations implemented at each step.}
    \label{fig:system_architecture}
\end{figure*}

The controller works in two stages: first, the controller performs the regulation of the embedding described in \eqref{eq:equival} by computing $\dot\phi_{d,i}$ in polar form, from which the feedback is calculated. We assume a fixed desired radius $r_d$ of the embedding. In stage two, this result is transformed back into Cartesian form and rotated by $\bm{\rho}_i(\phi)$ to generate the desired curve. 
Drawing from previous work in dynamic encirclement \cite{hafez-2015}, we can define the following controller:

\begin{equation}\label{eqn:angular1}
        \dot\phi_{d,i}
    =
    \frac{1}{3}
    \begin{bmatrix}
        2k_{\phi} & -k_{\phi} & -k_{\phi} & 3 
    \end{bmatrix}
    \begin{bmatrix}
    \phi_i \\
    \phi_j \\
    \phi_k \\
    \dot\phi_d 
    \end{bmatrix},
\end{equation}

\noindent where $k_\phi$ is a gain parameter, $\phi_i$ is the angle of the agent for which a control signal is to be generated, $\phi_j$ is the angle of the lagging agent, and $\phi_k$ is the angle of the leading agent. We then perform a transformation back to Cartesian coordinates as

\begin{equation}\label{eqn:xd}
    \hat{\bm{x}}_{d,i} 
    =
    \begin{bmatrix}
        r_{d}\cos(\phi_i)  \\
        r_{d}\sin(\phi_i) \\
        0
    \end{bmatrix}\text{ and }
    \hat{\bm{v}}_{d,i} 
    =
    \begin{bmatrix}
         0 \\
         0 \\
         \dot{\phi}_{d,i}
    \end{bmatrix}.
\end{equation}


Recalling \eqref{eq:rho}, the reference trajectory for agent $i$ is then computed as
\begin{equation} \label{eqn:xd2}
    \bm{x}_{d,i} = \bm\rho_i(\phi) * \hat{\bm{x}}_{d,i} *\bm\rho_i^{-1}(\phi),
\end{equation}
\begin{equation} \label{eqn:vd2}
    \bm{v}_{d,i} = (\bm\rho_i(\phi) * \hat{\bm{v}}_{d,i} *\bm\rho_i^{-1}(\phi))\times \bm{x}_i,
\end{equation}
\noindent for which we derive a linear control policy as
\begin{equation} \label{eq:cntrler}
    \bm{u_i} = k_x(\bm{x}_{d,i} - \bm{x}_{i}) + k_v(\bm{v}_{d,i}-\bm{v}_{i}),
\end{equation}
\noindent where $k_x$ and $k_v$ are gains selected for the desired performance and stability characteristics. 

Note that a system with dynamics \eqref{eq:dyn} controlled by \eqref{eq:cntrler} is asymptotically stable when $k_x > 0$ and $k_v^2 > 4k_x$. This can be seen from the reformulation of~\eqref{eq:dyn} in a state-space representation
\begin{equation}
    \begin{bmatrix}
        \dot{\bm{x}}_i \\
        \dot{\bm{v}}_i
    \end{bmatrix}
    =
    \begin{bmatrix}
        0 & 1 \\
        0 & 0
    \end{bmatrix}
    \begin{bmatrix}
        {\bm{x}}_i \\
        {\bm{v}}_i
    \end{bmatrix}
    +
    \begin{bmatrix}
        0 \\
        1
    \end{bmatrix}
    \bm{u}_i.
\end{equation}

Setting the reference state at the origin ($(\bm{x}_{d,i}, \bm{v}_{d,i}) = (0,0)$), then we reformulate \eqref{eq:cntrler} as
\begin{equation}
    \bm{u}_i
    = 
    -
    \begin{bmatrix}
    k_x &
    k_v
    \end{bmatrix}
    \begin{bmatrix}
    \bm{x}_i \\
    \bm{v}_i
    \end{bmatrix},
\end{equation}

\noindent which resolves to
\begin{equation}
    \begin{bmatrix}
        \dot{\bm{x}}_i \\
        \dot{\bm{v}}_i
    \end{bmatrix}
    =
    \underbrace
    {
    \begin{bmatrix}
        0 & 1 \\
        -k_x & -k_v
    \end{bmatrix}
    }_{A_x}
    \begin{bmatrix}
        {\bm{x}}_i \\
        {\bm{v}}_i
    \end{bmatrix},
\end{equation}

\noindent where $A_x$ is a new state matrix. Let us define a vector of eigenvalues $\bm{\lambda}$ and eigenvectors $\bm{\bm{\mu}}$ such that
\begin{equation}
    A_x \bm\mu = \bm{\lambda} \bm\mu
\end{equation}

\noindent where, given identity matrix $I$, $\bm{\lambda}$ is derived from the characteristic equation $\text{det}(A_x - \bm{\lambda}I) = 0$. Therefore, it follows that $A_x$ is asymptotically stable when all elements of $\bm{\lambda}$ have negative real parts \cite{7801025}. Assuming we do not desire oscillations (and hence, no imaginary parts), solving for $\bm{\lambda}$ we obtain:
\begin{equation}
    \bm{\lambda} = 
    \begin{bmatrix}
    -k_v/2 - \frac{\sqrt{k_v^2 - 4k_x}}{2} & \quad -k_v/2 + \frac{\sqrt{k_v^2 - 4k_x}}{2}
    \end{bmatrix}
\end{equation}

\noindent which are negative for $k_x > 0$ and $k_v^2 > 4k_x$.

\section{Experimental Methodology}
\label{sec:methodology}



\begin{figure}
    \centering
    \includegraphics[width=0.75\columnwidth]{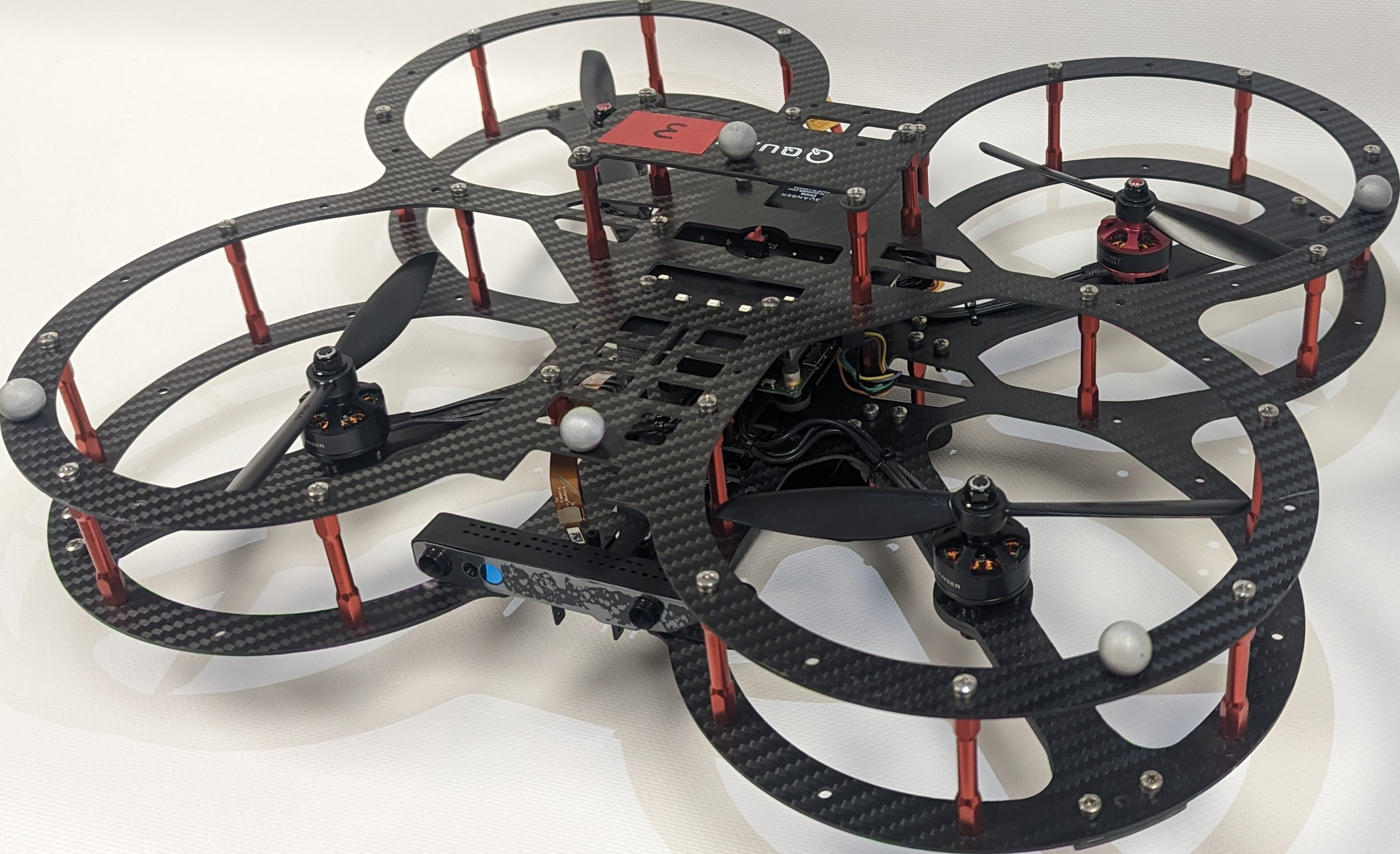}
    \caption{Quanser QDrone used for physical experiments.}
    \label{fig:qdrone}
\end{figure}

In this section, we implement the system proposed in Sections \ref{sec:problem} and \ref{sec:control} on three real quadcopters in order to validate the design experimentally. 
Fig.~\ref{fig:system_architecture} illustrates the three parts of our system (the trajectory generation with circle embedding, the control, and the UAV dynamics) and how they interact. The platforms used for the experiments were the Quanser QDrones shown in Fig.~\ref{fig:qdrone}. The low-level controllers, which receive acceleration as input, were developed by Quanser and were modified only to receive the signals computed in Section \ref{sec:control}. The entire QDrone platform, which includes low-level controllers, is depicted in Fig.~\ref{fig:system_architecture} as a green block. The controllers were implemented in Matlab\texttrademark~and ported to each vehicle separately. Therefore, each vehicle generated its own control signal in a decentralized manner based on its onboard observations and independently of the other vehicles. 

\begin{figure}
    \centering
    \includegraphics[width=0.9\columnwidth]{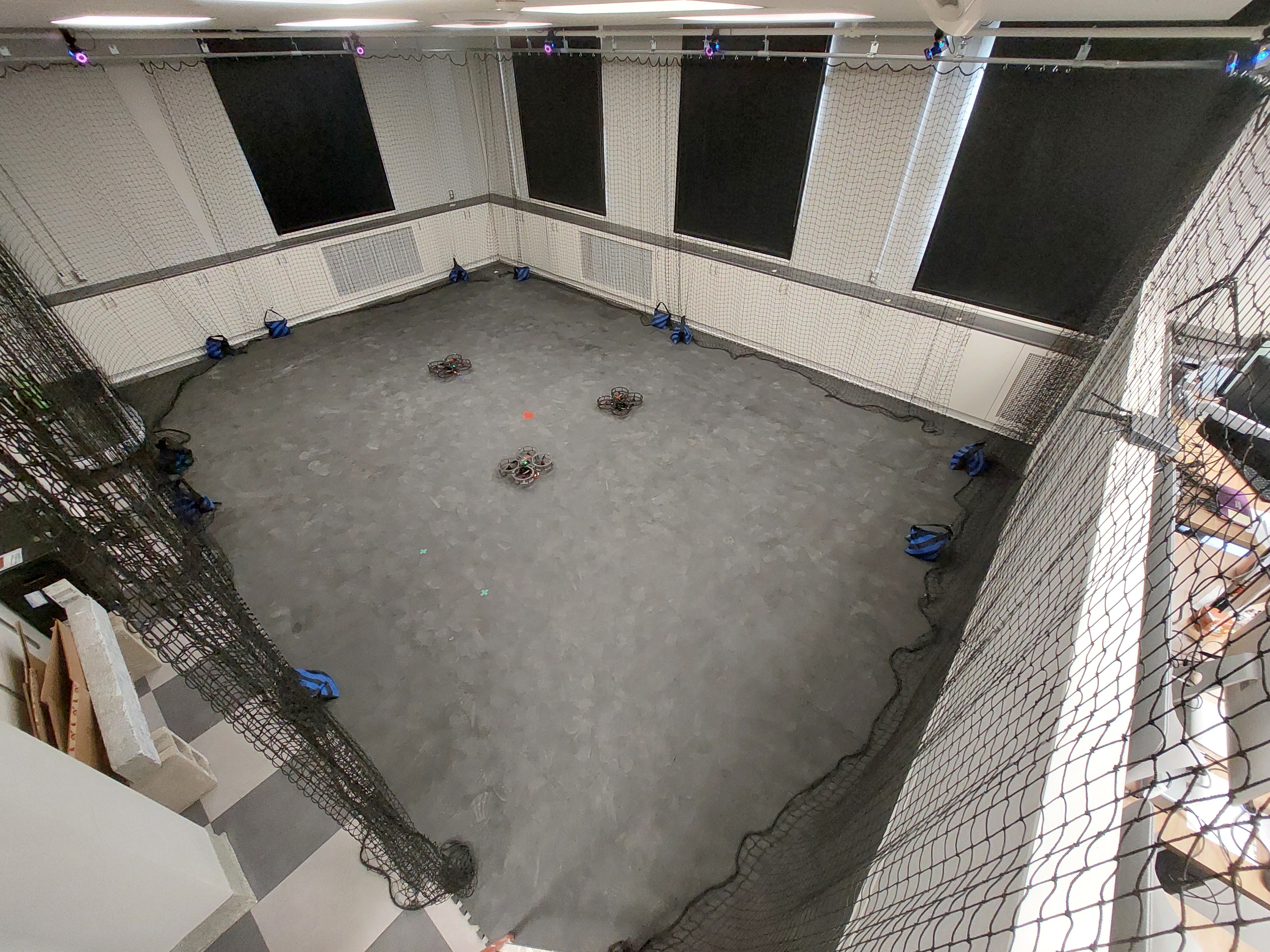}
    \caption{Test area for autonomous flight.}
    \label{fig:testarea}
\end{figure}

The UAVs flew in an indoor environment of dimensions $5 \times 7.5 \times 3$~m (width, length, and height respectively), shown in Fig.~\ref{fig:testarea}. 
Position and orientation of each UAV were monitored using a Vicon Camera System\footnote{\url{www.vicon.com}}, with $10$ cameras, communicating with a ground station that collects the camera images, calculates UAVs' states, and sends them to the vehicles, at a frequency of $100$~Hz. Note that the indoor camera setup was used for experiments given that it was readily available to provide safe UAV flight. However, the solution proposed in this work is inherently decentralized, as evidenced in Section~\ref{sec:control}, and does not rely on a central positioning system.  Moreover, considering a particular vehicle $i \in \{1,2,3\}$, the states $(\bm{x}_i,\bm{v}_i)$ are fed back for the calculation of the control signal. Finally, vehicle $i$ also receives the polar coordinates (angles) of its neighbours (lagging and leading vehicles), to enable computation of equation~\eqref{eqn:angular1}. 

\section{Results}
\label{sec:results}

To validate the proposed approach, we first provide a simulation of the system in Section~\ref{sec:simulation} and then show the experiments on real platforms in Section~\ref{sec:experiment}. In both cases, the UAVs are configured to follow a dumbbell curve for approximately $40~\text{s}$, with $\dot{\phi_d}=0.2$, $k_{\phi} = 2.0$, and $r_d=1.5$~m. The dynamics assumed in the controller were the double integrator model in~\eqref{eq:dyn}. The control architecture shown in Fig.~\ref{fig:system_architecture} was implemented in Python for the simulation and, as mentioned in Section~\ref{sec:methodology}, in Matlab for the experiment (with only an executable being transferred to each UAV at initialization).

\subsection{Simulation} \label{sec:simulation}
\begin{figure}
\centering
\includegraphics[width=\columnwidth]{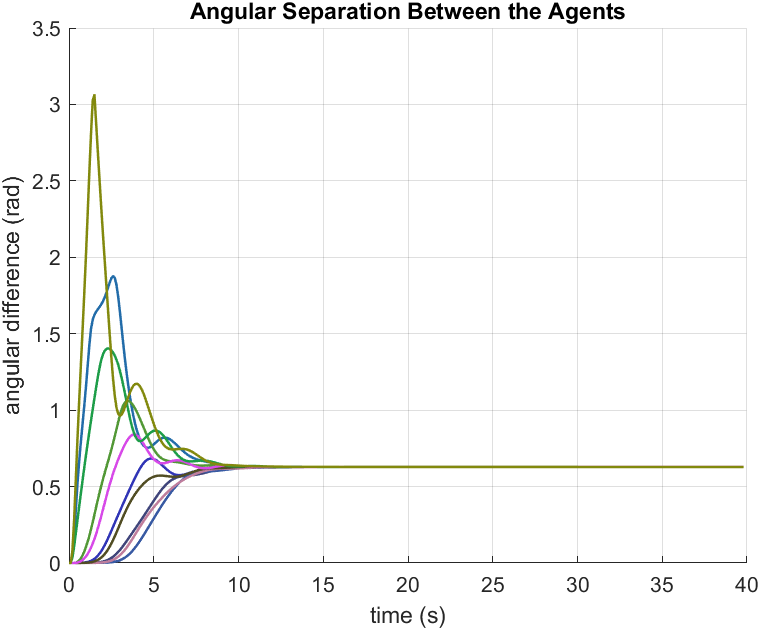}
\caption{Angular separation between the UAVs throughout the Dumbbell trajectory}
\label{fig:phase-sep-sim}
\end{figure}

\begin{figure}
\centering
\includegraphics[width=\columnwidth]{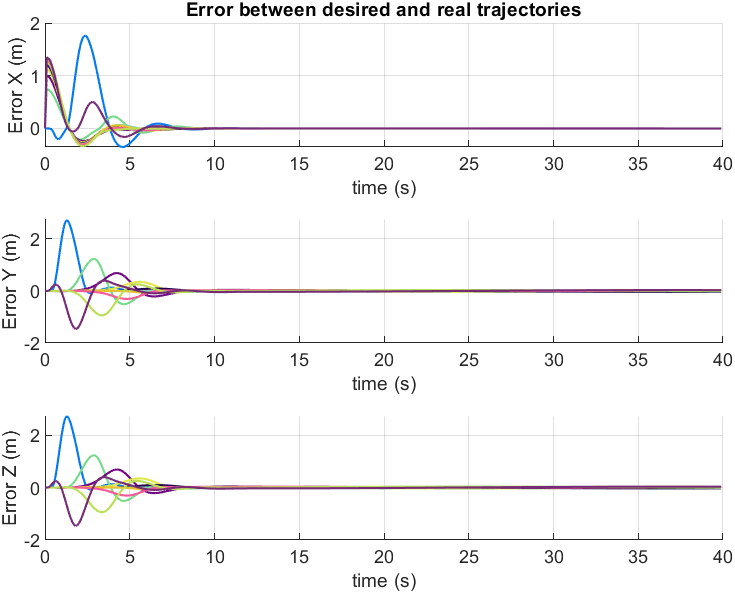}
\caption{Error between X, Y and Z desired and real coordinates of the UAVs, following a Dumbbell curve, as a function of time.}
\label{fig:error1-xyz-sim}
\end{figure}

To demonstrate that our approach can be executed with an arbitrary number of agents in the system, a simulation trial was performed with $10$ agents. In Fig.~\ref{fig:phase-sep-sim}, note that the agents achieved a uniform separation of $2\pi/10$ rad, which is the expected value, after approximately $10$~s. The errors between the desired and the simulated trajectories, displayed in Fig.~\ref{fig:error1-xyz-sim}, for all the agents, vanish in all the axes also after $10$~s.

\subsection{Experiment}
\label{sec:experiment}

\begin{figure}
\centering
\includegraphics[width=\columnwidth]{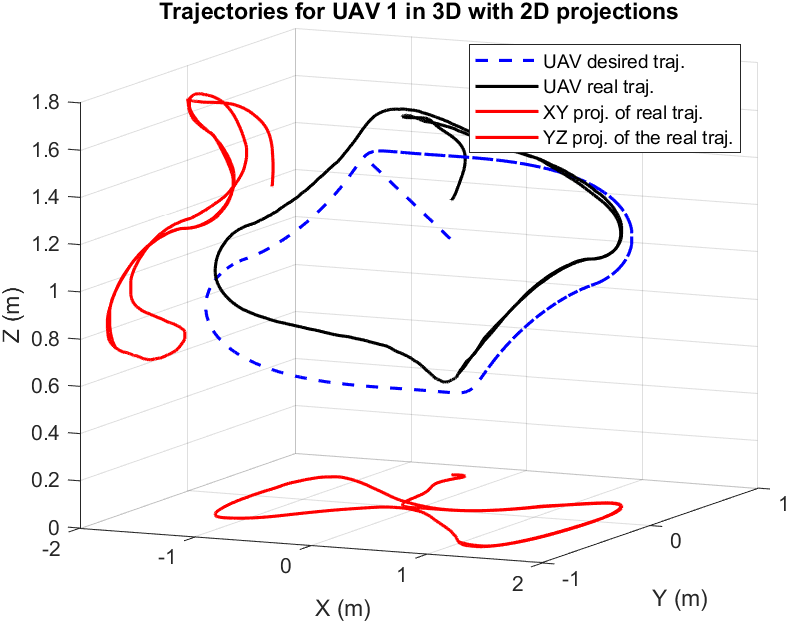}
\caption{Trajectories of UAV 1 in 3D, following a Dumbbell curve, and its projection on the X-Y and Y-Z planes.}
\label{fig:drone1-proj}
\end{figure}

\begin{figure}
\centering
\includegraphics[width=\columnwidth]{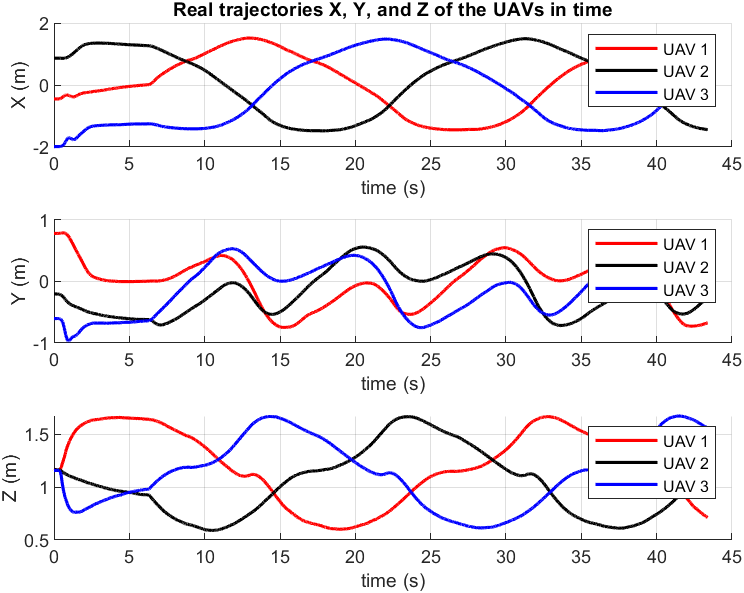}
\caption{X, Y and Z trajectories executed by the UAVs, following a Dumbbell curve, as a function of time.}
\label{fig:drones-xyz}
\end{figure}

\begin{figure}
\centering
\includegraphics[width=\columnwidth]{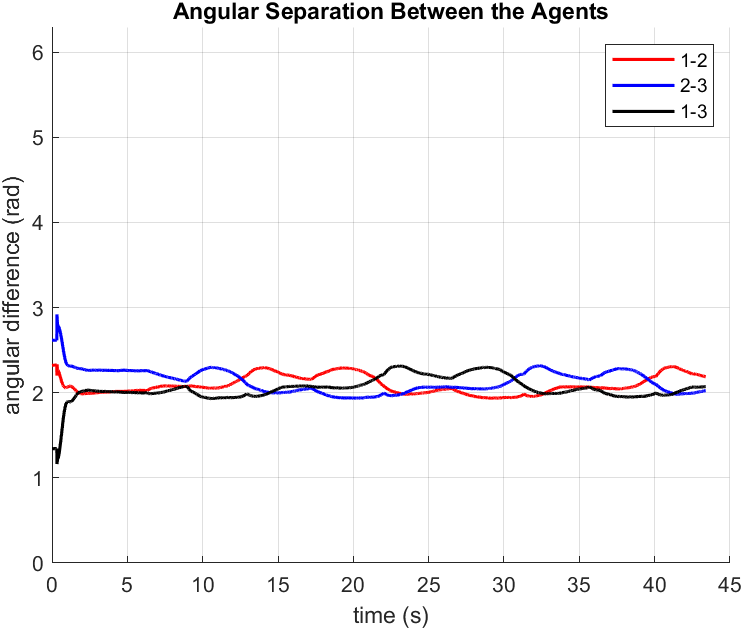}
\caption{Angular separation between the UAVs throughout the Dumbbell trajectory}
\label{fig:phase-sep}
\end{figure}

Fig.~\ref{fig:drone1-proj} shows the desired and real trajectories of vehicle 1, along with its projection on the $X-Y$ and $Y-Z$ planes. 
Fig.~\ref{fig:drones-xyz} displays the position coordinates of the three vehicles during the experiment. Note that the vehicle positions behave as periodic functions in time, with a phase shift shown in Fig.~\ref{fig:phase-sep}. This figure illustrates the desired even separation between each vehicle. The difference in angle oscillates near $2\pi/3~\text{rad}$, meaning that the controller designed to operate within the embedding was able to keep the vehicles uniformly spaced (in polar coordinates, $\phi$), as expected. 

\begin{figure}
\centering
\includegraphics[width=\columnwidth]{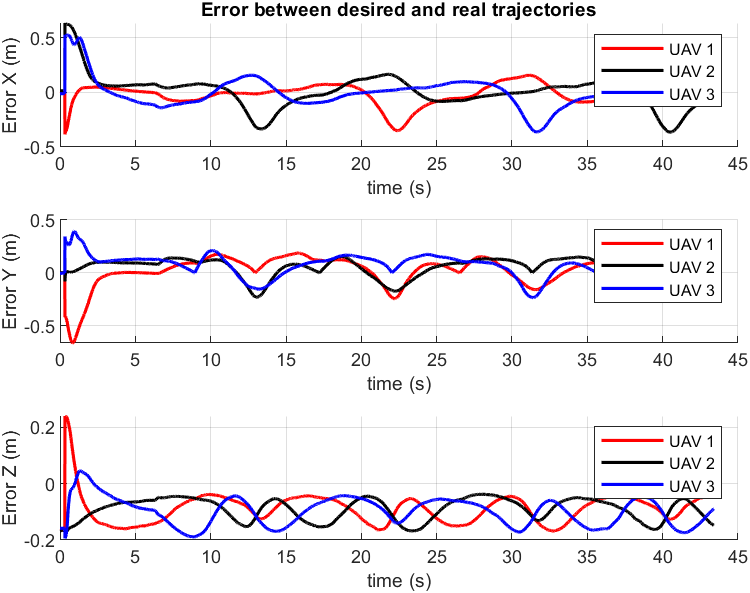}
\caption{Error between X, Y and Z desired and real coordinates of the UAVs, following a Dumbbell curve, as a function of time.}
\label{fig:error1-xyz}
\end{figure}

The error between the desired and real trajectories of all the vehicles is shown in Fig.~\ref{fig:error1-xyz}. An increase in the $x$ and $y$ error curves can be noticed at $13$~s, $22$~s, and $ 31$~s.
The position of UAVs (Fig.~\ref{fig:drones-xyz}) suggests that such errors are caused by air turbulence due to UAVs moving close to one being on top of another. However, these disturbances did not compromise the stability or navigation performance, which is evidence of the robustness of the proposed technique.  

Additionally, Table~\ref{tab:rms-dumb} shows the Root Mean Squared Error (RMSE) between the desired and real coordinates of all the drones. The RMSE is about $0.1$~m for the three axes, which represent less than $10\%$ of the trajectory's radius (which was set to $1.5$~m) and around one quarter of the diameter of the vehicle ($0.5$~m). This average error and the angular separation of the UAVs not remaining constant at $2\pi/3~\text{rad}$ (as in simulation) were caused by disturbances in the system and unmodelled dynamics not considered in the development of the controller. However, they demonstrate the satisfactory performance of our approach even under adverse conditions.
\begin{table}[]
\centering
\caption{RMSE between the desired and the real x, y, and z coordinates of the three drones, following a Dumbbell curve.}
  \begin{tabular}{c|c|c|c}
Drone & RMSE X ($m$) & RMSE Y ($m$) & RMSE Z ($m$) \\ \hline
1       & 0.093   & 0.110   & 0.101   \\ \hline
2       & 0.120   & 0.109   & 0.096   \\ \hline
3       & 0.110   & 0.120   & 0.113  
\end{tabular}  
\label{tab:rms-dumb}
\end{table}

\section{Conclusion}
\label{sec:conclusion}

This work introduces a novel, decentralized system for generating self-organizing curve trajectories in swarms of agents. The core element of this approach is its use of a quaternion-based stabilizing embedding that is topologically equivalent to the actual trajectory. The formulation of this embedding permits the use of well-established linear control policies in an application where this would be otherwise infeasible. The overall structure of the swarm is emergent, derived from local observations, and without the use of a global planner.
By employing linear control policies on these local observations, our approach is extendable to systems involving a larger number of agents. The development and analysis of the control architecture is based on double integrator dynamics, which we demonstrate is robust to the unmodelled low-level dynamics of real quadcopters. 

Navigation in real-world conditions is characterized by complex dynamics and environmental challenges, such as air turbulence. Therefore, practical experiments highlight the generality and robustness of the proposed embedding method.  Notably, our method's adaptability extends beyond quadcopters, making it suitable for implementation in various UAV configurations. 

Finally, future research will explore the investigation of how system constraints, such as maximum speed or acceleration, influence the embedding.


\balance








\bibliographystyle{IEEEtran}
\bibliography{references,Travis}

\end{document}